\documentclass[12pt,preprint]{aastex}
\usepackage{amssymb}
\usepackage{amsmath}
\usepackage{epstopdf}
\usepackage{natbib}
\usepackage[colorlinks, citecolor=blue]{hyperref}
\usepackage{hyperref}
\usepackage{graphicx}
\usepackage{subfigure}
\usepackage[all]{hypcap}

\begin{document}

\title{$\rm\textbf{TeV-PeV}$ Neutrino Oscillation of Low-luminosity Gamma-ray Bursts}
\author{D. Xiao\altaffilmark{1,2} and Z. G. Dai\altaffilmark{1,2}}
\affil{\altaffilmark{1}School of Astronomy and Space Science, Nanjing University, Nanjing 210093, China; dzg@nju.edu.cn}
\affil{\altaffilmark{2}Key Laboratory of Modern Astronomy and Astrophysics (Nanjing University), Ministry of Education, China}

\begin{abstract}

There is a sign that long-duration gamma-ray bursts (GRBs) originate from the core collapse of massive stars. During a jet puncturing through the progenitor envelope, high energy neutrinos can be produced by the reverse shock formed at the jet head. It is suggested that low-luminosity GRBs (LL-GRBs) are possible candidates of this high energy neutrino precursor up to $\sim {\rm PeV}$. Before leaving the progenitor, these high energy neutrinos must oscillate from one flavor to another with matter effect in the envelope. Under the assumption of a power-law stellar envelope density profile $\rho \propto r^{-\alpha}$ with an index $\alpha$, we study the properties of ${\rm TeV-PeV}$ neutrino oscillation. We find that adiabatic conversion is violated for these neutrinos so we do certain calibration of level crossing effect. The resonance condition is reached for different energies at different radii. We notice that the effective mixing angles in matter for ${\rm PeV}$ neutrinos are close to zero so the transition probabilities from one flavor to another are almost invariant for ${\rm PeV}$ neutrinos. We plot all the transition probabilities versus energy of ${\rm TeV-PeV}$ neutrinos from the birth place to the surface of the progenitor. With an initial flavor ratio $\phi_{\nu_e}^0:\phi_{\nu_\mu}^0:\phi_{\nu_\tau}^0=1:2:0$, we plot how the flavor ratio evolves with energy and distance when neutrinos are still in the envelope, and further get the ratio when they reach the Earth. For ${\rm PeV}$ neutrinos, the ratio is always $\phi_{\nu_e}:\phi_{\nu_\mu}:\phi_{\nu_\tau}\simeq0.30:0.37:0.33$ on Earth. In addition, we discuss the dependence of the flavor ratio on energy and $\alpha$ and get a pretty good result. This dependence may provide a promising probe of the progenitor structure.

\end{abstract}


\section{Introduction}

The idea that GRBs can serve as the sources of high energy neutrinos has long been discussed \citep{wax97, wax00, dai01, li02, der03, raz13, vie95, mes00, gue04, mur06, ree05, mur08, wan09, gao12, mes01, enb09, mur13, pru03, raz03, and05, hor08}. Before breaking out, a relativistic jet punctures through the stellar envelope and transports energy to electrons and protons via shock acceleration \citep{zha04, pir05, mes06, woo93, mac99}. Accelerated electrons dominate radiation by synchrotron or inverse Compton mechanism while accelerated protons produce neutrinos by proton-proton collision and photopion process \citep{wax97, wax99, rac98, alv00, bah01, gue03, mura06, bec08}. This neutrino signal is prior to the main burst, as a precursor. However, it has been argued that this high energy neutrino precursor of energy ranging from ${\rm TeV}$ up to ${\rm PeV}$ cannot be produced for a typical GRB \citep{lev08, kat10, mur13}. The reason is that the reverse shock occurring at the interface of the jet head would be radiation mediated, resulting in an inefficient shock acceleration. Nevertheless, a matter is different for LL-GRBs \citep{mur13, xia14}. Due to its low power, the Thomson optical depth is low even inside a star so that efficient shock acceleration would be expected. We assume the same LL-GRB as our previous work \citep{xia14}, in which we have shown that our LL-GRB is responsible for ${\rm TeV-PeV}$ neutrinos. Further in this paper we focus on the oscillation properties of these high energy neutrinos.
\par
Neutrino oscillation in matter has been studied for a long while. The resonant conversion of neutrinos from one flavor to another with matter effect was first discussed in solar neutrino problem \citep{che85}. While propagating in a medium, $\nu_e$ interacts via neutral current (NC) and charged current (CC), whereas $\nu_\mu$ and $\nu_\tau$ interacts only via NC. This mechanism is called the Mikheyev-Smirnov-Wolfenstein (MSW) effect \citep{mik85, wol78}. As we know, solar neutrinos are relative low in energy ($\leq {100\rm MeV}$), as is the same for supernova neutrinos, of which the research is becoming more and more attractive \citep{lun13, sch12}. In the meantime, osillation of GRB neutrinos is yet less understood. With energy up to ${\rm PeV}$, the oscillation properties of GRB neutrinos are different compared to ${\rm MeV}$ neutrinos. \citet{kas05} discussed the electromagnetic and adiabatic energy losses of $\pi'$s and $\mu'$s which would modify the flavor ratio produced by GRBs and concluded that the flavor ratio on Earth $\phi_{\nu_e}:\phi_{\nu_\mu}:\phi_{\nu_\tau}$ is $1:1:1$ at low energy to $1:1.8:1.8$ at high energy with a transition energy around $100 \rm TeV$. \citet{men07} discussed high-energy neutrinos produced in optically thick astrophysical objects in the energy range $0.1-100\rm TeV$. They studied in detail the shower-to-muon track ratio $R$ and discussed its variation with source properties and neutrino oscillation parameters.  \citet{raz10} also presented a detailed and comprehensive study of flavor conversion of neutrinos from hidden sources (jets) but their results differ from the results of \citet{men07}. \citet{sah10} showed that the resonant oscillation could take place within the inner high density region of the choked jet progenitor and the final flavor ratio detected on Earth is further modified to either $1:1.095:1.095$ for the large mixing angle solution to the solar neutrino data, or $1:1.3:1.3$ for maximal mixing among the muon and tau neutrinos in vacuum. \citet{oli13} studied the three flavor neutrino oscillation of choked GRBs in GeV-TeV energy range for three different presupernova star models. They found that for neutrino energies below $\leq10\rm TeV$ the flux ratio did not amount to $1:1:1$, whereas for higher energy neutrinos it did. Recently, \citet{fra14} carried out an analysis of resonance conditions for $\rm GeV-PeV$ neutrinos created in internal shocks at different places in the star, estimating the flavor ratios on Earth.
\par
The main difference of our paper with previous works is at the starting point that we take a LL-GRB as the source of high energy neutrinos and hence we focus on higher energy from ${\rm TeV}$ to ${\rm PeV}$. We find that the mixing angles in matter for ${\rm PeV}$ neutrinos are close to zero so the transition probabilities from one flavor to another are almost invariant, which are different with ${\rm MeV-TeV}$ neutrinos. Thus we get a constant ratio of $\phi_{\nu_e}:\phi_{\nu_\mu}:\phi_{\nu_\tau}\simeq0.30:0.37:0.33$ for $\rm PeV$ neutrinos on Earth. Moreover, we discuss the dependence of the flavor ratio on the index $\alpha$ and neutrino energy, providing a promising way to probe the GRB progenitor structure through a neutrino precursor signal in the future.
\par
This paper is organised as follows. We present all results in section 2. Subsection 2.1 is about the density profile of the envelope and subsection 2.2 is about adiabatic violation for neutrinos above $\rm TeV$. In subsection 2.3 we discuss the three neutrino mixing both in the envelope and in vacuum, and then the dependence on envelope density profile index is exhibited in subsection 2.4. We finish with discussions and conclusions in section 3.

\section{Neutrino Mixing}
\subsection{Density Profile of the Envelope}
In this subsection, we take $\alpha=2$ as our premise, and discuss the dependence on $\alpha$ later in subsection 2.4.
\par
We assume a power-law envelope density profile $\rho(r) =A r^{-\alpha}$, where $A=(3-\alpha)M_{\rm He}/(4\pi R^{3-\alpha})$ and $2\leq \alpha < 3$ with $M_{\rm He}$ and $R$ being the mass and radius of the helium envelope. For a helium core of mass $M_{\rm He}=2M_\odot$ and radius $R=4\times10^{11}\,$cm, the ambient envelope density can be expressed as $\rho(r)=7.96\times 10^{20}r^{-2}\;{\rm g\,cm}^{-3}$. The number density of electrons in the envelope is $N_e(r)=\frac{\rho(r)}{4m_p}Y_e$, where the number of electrons per nucleon $Y_e$ needs to be obtained.
\par
The Saha's equation reads
$$\log\frac{N_{r+1}}{N_r}=\log\frac{2u_{r+1}(T)}{u_r(T)}+\frac{5}{2}\log T-\frac{5040}{T}\chi_r-\log P_e-0.48, \eqno (1)$$
where $N_r, u_r$, and $\chi_r$ stand for the number density, partition function and ionization energy of $r_{th}$ ionization ions, respectively. $T$ is the temperature and $P_e\equiv N_ekT$ is the electron pressure.
For our pure helium envelope, we can get
$$\frac{N(\rm He^{2+})}{N(\rm He^+)}=1.65\times10^{-11},\,\,\frac{N(\rm He^+)}{N(\rm He)}=0.69, \eqno (2)$$
if we adopt typical values as $T=15000\mathrm K, \chi_0=24.58\rm eV, \chi_1=54.4\rm eV$. We can see that the 2nd ionization of helium is negligible so that $Y_e=\frac{N(\rm He^+)}{N(\mathrm {He})+N(\rm He^+)}\simeq0.408$.
\par
The effective potential of neutrinos can be expressed as
$$ V_{\rm eff}=\sqrt{2}G_FN_e,  \eqno (3)$$
where $G_F$ is the Fermi coupling constant.
\subsection{Adiabatic Conversion Violation}
\begin{figure}[htbp]
\centering
\includegraphics[width=0.7\textwidth]{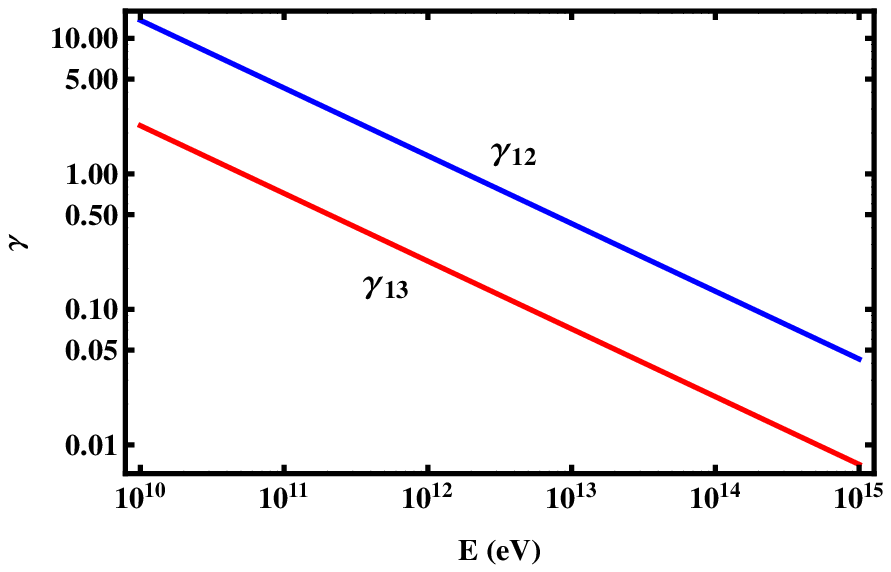}
\caption{Adiabatic parameters versus neutrino energy of our LL-GRB progenitor. The red line represents adiabatic parameter for level crossing between 1st and 3rd energy eigenstates, while the blue line is that for  level crossing between 1st and 2nd energy eigenstates. \label{Figure 1}}
\end{figure}

Before we start to consider the neutrino oscillation, we need to check whether the adiabatic approximation is valid. The adiabatic parameter $\gamma$ is defined as $$\gamma\equiv \frac{\delta m^2}{2E}\sin{2\theta}\tan{2\theta}\frac{1}{\vert{\frac{d\ln N_e}{dr}}\vert_{\rm res}}, \eqno (4)$$
where $\delta m^2$ is the mass square difference between the neutrino mass eigenstates, $E$ is the neutrino energy and $\theta$ is the mixing angle. The subscript $``\mathrm {res} "$ represents the place at which resonance happens. We can easily see that $\gamma$ is in proportion to $1/E$ and the adiabatic approximation requirement $\gamma\gg1$ is usually fulfilled for neutrinos with relative low energy of $\leq100\rm MeV$, such as solar neutrinos and supernovae neutrinos. However, in this paper we focus on high energy from $\rm TeV$ to $\rm PeV$ and we find that adiabatic conversion is not applicable now. We plot the adiabatic parameter versus neutrino energy in Figure~\ref{Figure 1} and it is obvious that $\gamma\gg1$ is violated for high energy neutrinos. Only $\gamma_{12}$ and $\gamma_{13}$ are needed because there are at most two level crossing for neutrinos in the three flavor case \citep{dig00, yas14}. The vacuum oscillation parameters we adopt are $\delta m_{12}^2=7.54\times10^{-5}\mathrm {eV}^2,\, \delta m_{23}^2=2.43\times10^{-3}\rm eV^2,\, \sin^2{\theta_{12}}=0.307, \, \sin^2{\theta_{13}}=2.41\times10^{-2}, \,\sin^2{\theta_{23}}=3.86\times10^{-1}$, CP violation phase $\delta=1.08\pi$ and normal mass hierarchy is assumed \citep{fog12}.
\par
\begin{figure}[htbp]
\centering
\includegraphics[width=0.7\textwidth]{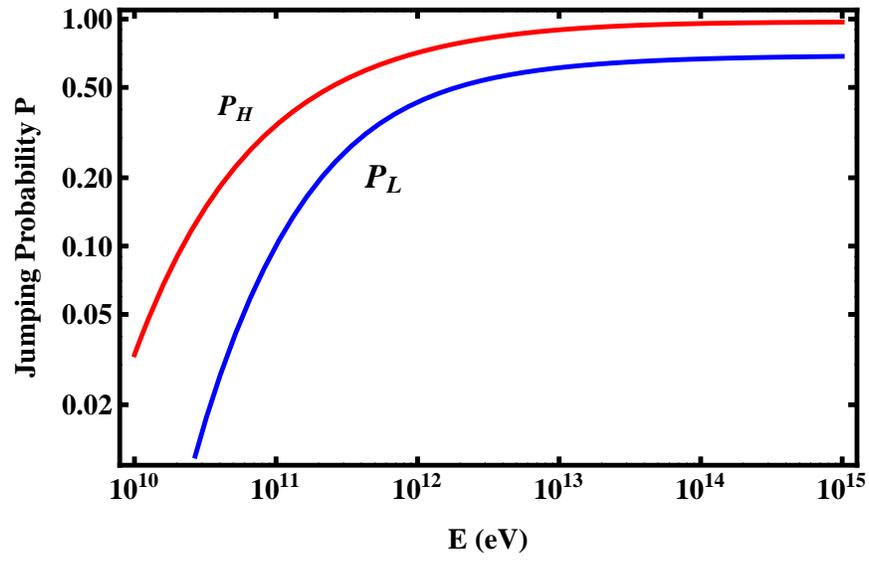}
\caption{Jumping probabilities versus neutrino energy according to adiabatic parameters. The red line represents the jumping probability between 1st and 3rd energy eigenstates, while the blue line is that for the jumping probability between 1st and 2nd energy eigenstates. \label{Figure 2}}
\end{figure}

For the reason above, we are obliged to do certain calibration of level crossing effect. The jumping probability is approximately computed by the WKB method \citep{yas14, kuo89}:
$$P=\frac{\exp[-\frac{\pi}{2}\gamma F]-\exp[-\frac{\pi}{2}\gamma \frac{F}{\sin^2\theta}]}{1-\exp[-\frac{\pi}{2}\gamma \frac{F}{\sin^2\theta}]},  \eqno (5)$$
where $F$ is a factor depending on the density profile. Then  in Figure~\ref{Figure 2} we plot $P_H$ and $P_L$ versus neutrino energy, representing the jumping probability from energy eigenstate $\nu_{1m}$ to $\nu_{3m}$ and from $\nu_{1m}$ to $\nu_{2m}$ respectively.
\subsection{Three-neutrino Mixing}
\subsubsection{Neutrino Oscillation in the Envelope}
The evolution equation for neutrinos in matter is given by
$$i\frac{d\Psi}{dt}=[UH_0U^\dag+V_{\rm eff}]\Psi,  \eqno (6)$$
where $H_0=\frac{1}{2E}\mathrm{diag}(-\delta m_{21}^2,\,0,\,\delta m_{32}^2)$ and $\Psi^T\equiv(\nu_e,\nu_\mu,\nu_\tau)$ is the flavor eigenstate \citep{fra14}. $U$ is the three neutrino mixing matrix,
\begin{displaymath}
U=\left( \begin{array}{ccc}
c_{12}c_{13} & s_{12}c_{13} & s_{13}e^{-i\delta}\\
-s_{12}c_{23}-c_{12}s_{23}s_{13}e^{i\delta} & c_{12}c_{23}-s_{12}s_{23}s_{13}e^{i\delta} & s_{23}c_{13}\\
s_{12}s_{23}-c_{12}c_{23}s_{13}e^{i\delta} & -c_{12}s_{23}-s_{12}c_{23}s_{13}e^{i\delta} & c_{23}c_{13}
\end{array} \right),
\eqno (7)
\end{displaymath}
where $s_{ij}\equiv\sin{\theta_{ij}},\,c_{ij}\equiv\cos{\theta_{ij}}$.
\par
\begin{figure}
\centering
\subfigure[Effective mixing angles versus neutrino energy at born site $r_0=1\times10^{11}\rm cm$]{
\begin{minipage}[b]{0.45\textwidth}
\includegraphics[width=1\textwidth]{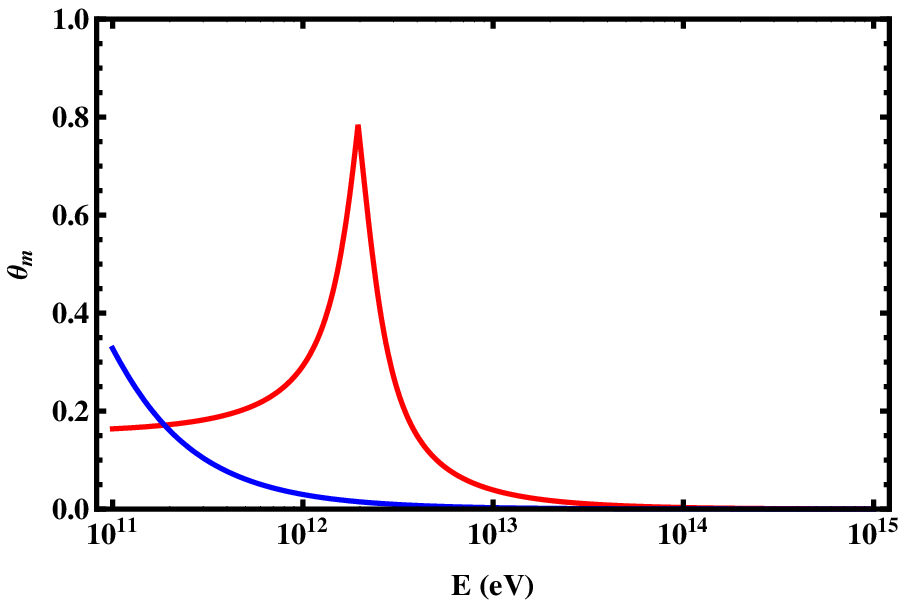}
\label{fig3:subfig:a}
\end{minipage}
}
\subfigure[Effective mixing angles versus neutrino energy at progenitor surface $R=4\times10^{11}\rm cm$]{
\begin{minipage}[b]{0.45\textwidth}
\includegraphics[width=1\textwidth]{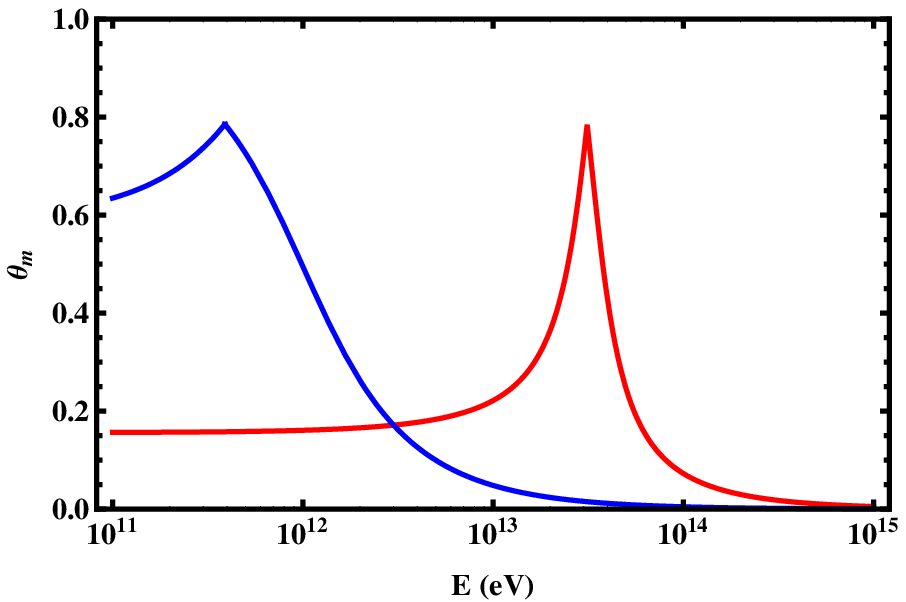}
\label{fig3:subfig:b}
\end{minipage}
}
\subfigure[Effective mixing angles evolve versus radius for $1\rm TeV$ neutrinos]{
\begin{minipage}[b]{0.45\textwidth}
\includegraphics[width=1\textwidth]{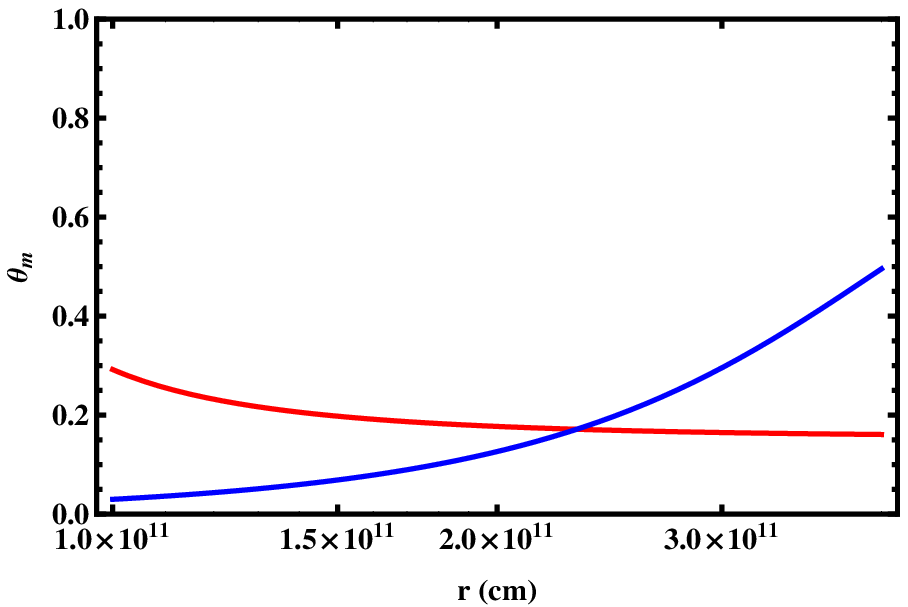}
\label{fig3:subfig:c}
\end{minipage}
}
\caption{Effective mixing angles in matter for neutrinos with different energies at different radii. In all three subfigures above, the red line represents $\theta_{13,m}$ and the blue line is $\theta_{12,m}$. All angles are measured in radians. \label{Figure 3}}
\end{figure}

Neutrino mixing angles in matter can be expressed as \citep{fra14, yas14}
$$\sin{2\theta_{13,m}}=\frac{\sin{2\theta_{13}}}{\sqrt{(\cos{2\theta_{13}}-2EV_{\mathrm {eff}}/\delta m_{13}^2)^2+(\sin{2\theta_{13}})^2}},$$
$$\sin{2\theta_{12,m}}=\frac{\sin{2\theta_{12}}}{\sqrt{(\cos{2\theta_{12}}-2EV_{\mathrm {eff}}c_{13}^2/\delta m_{12}^2)^2+(\sin{2\theta_{12}})^2}}, \eqno (8)$$
The effective mixing angles $\theta_{13,m},\,\theta_{12,m}$ become maximum $\pi/4$ as resonance conditions $\cos{2\theta_{13}}=2EV_{\mathrm {eff}}/\delta m_{13}^2$ and $\cos{2\theta_{12}}=2EV_{\mathrm {eff}}c_{13}^2/\delta m_{12}^2$ are fulfilled respectively. We can see that $\theta_{13,m},\,\theta_{12,m}$ are functions of neutrino energy and radius since $V_{\mathrm {eff}}=V_{\mathrm {eff}}(r)$. We plot $\theta_{13,m},\theta_{12,m}$ versus energy at radius $r_0=1.0\times10^{11}\rm cm$ in Figure~\ref{fig3:subfig:a} and find that $\theta_{13,m}$ reaches maximum for $\sim 2\rm TeV$ neutrinos. Here the radius $r_0$ is treated as the born site of high energy neutrinos. The reason is that the Thomson optical depth is $\tau_T=n\sigma_T l= \frac{\rho_j}{m_p}\sigma_T\frac{r}{\Gamma_j} \simeq0.158(\frac{L_j}{10^{42}{\rm erg\,s}^{-1}})^{\frac{2}{3}}(\frac{\theta_0}{0.02})^{\frac{29}{15}}(\frac{M_{\rm He}}{2M_\odot})^{\frac{1}{3}}(\frac{r}{4\times10^{11}\rm cm})^{-\frac{4}{3}}$ and at $r=r_0$ we have $\tau_T\simeq1$, thus ensuring the efficient shock acceleration \citep{xia14}. We also plot the effective mixing angles versus energy at progenitor surface $R=4\times10^{11}\rm cm$ in Figure~\ref{fig3:subfig:b} and how they evolve with propagation distance for $1\rm TeV$ neutrinos in Figure~\ref{fig3:subfig:c} respectively. In addition, we kindly find that effective mixing angles tend to be constant zero for $\rm PeV$ neutrinos, which is the main reason for constant flavor ratio and will be shown later.
\par
The transition probability from one flavor to another after level crossing calibration can be expressed as \citep{yas14}
\begin{displaymath}
\begin{split}
P(\nu_\alpha \to \nu_\beta)=\setlength\arraycolsep{2pt}\left(\begin{array}{ccc}\vert U_{\beta1,m}\vert^2 & \vert U_{\beta2,m}\vert^2 & \vert U_{\beta3,m}\vert^2
\end{array} \right)\left( \begin{array}{ccc}1-P_L & P_L & 0\\P_L & 1-P_L & 0\\0 &0&1\end{array} \right) \\ \times \left( \begin{array}{ccc}1-P_H & 0 & P_H\\0 & 1 & 0\\P_H &0&1-P_H\end{array} \right)\left( \begin{array}{ccc}\vert U_{\alpha1,m}\vert^2 \\ \vert U_{\alpha2,m}\vert^2 \\ \vert U_{\alpha3,m}\vert^2\end{array} \right),
\end{split}
\end{displaymath}$$\eqno (9)$$
where $\alpha,\beta\to e,\mu,\tau$ and $U_{\alpha i,m},U_{\beta j,m}$ are mixing matrix elements in matter.
\par
\begin{figure}
\centering
\subfigure[Transition probabilities versus neutrino energy at born site $r_0=1\times10^{11}\rm cm$]{
\begin{minipage}[b]{0.45\textwidth}
\includegraphics[width=1\textwidth]{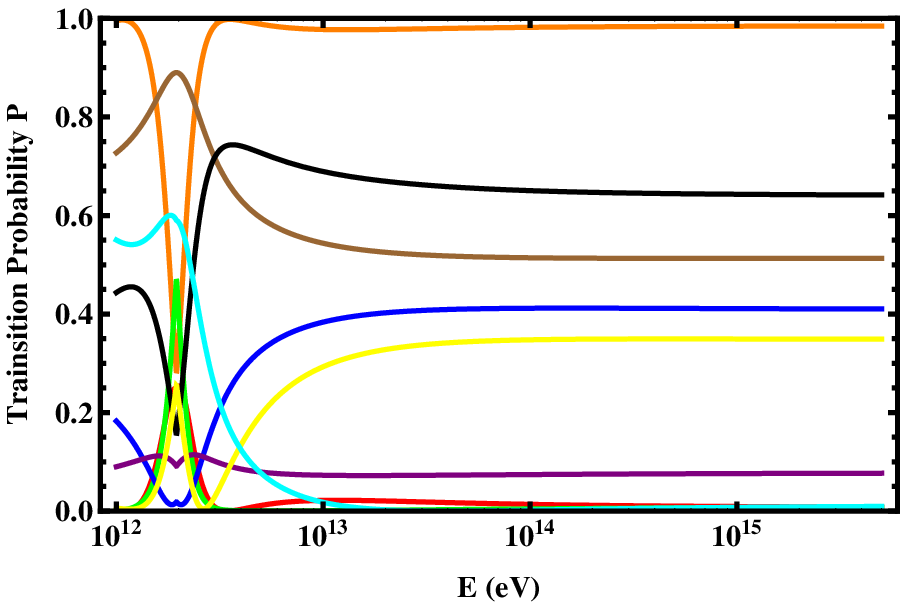}
\label{fig4:subfig:a}
\end{minipage}
}
\subfigure[Transition probabilities versus neutrino energy at progenitor surface $R=4\times10^{11}\rm cm$]{
\begin{minipage}[b]{0.45\textwidth}
\includegraphics[width=1\textwidth]{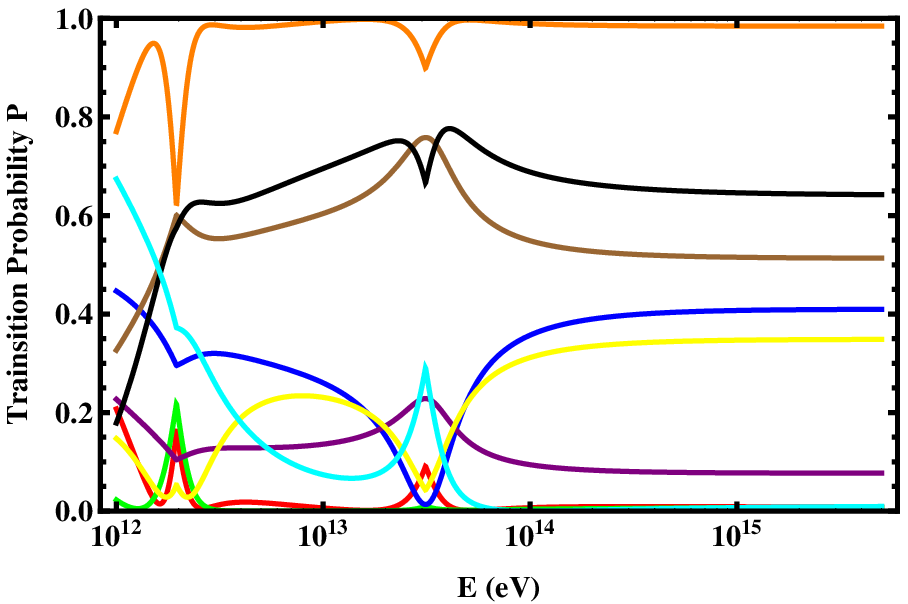}
\label{fig4:subfig:b}
\end{minipage}
}
\subfigure[Transition probabilities evolve versus radius for $1\rm TeV$ neutrinos]{
\begin{minipage}[b]{0.45\textwidth}
\includegraphics[width=1\textwidth]{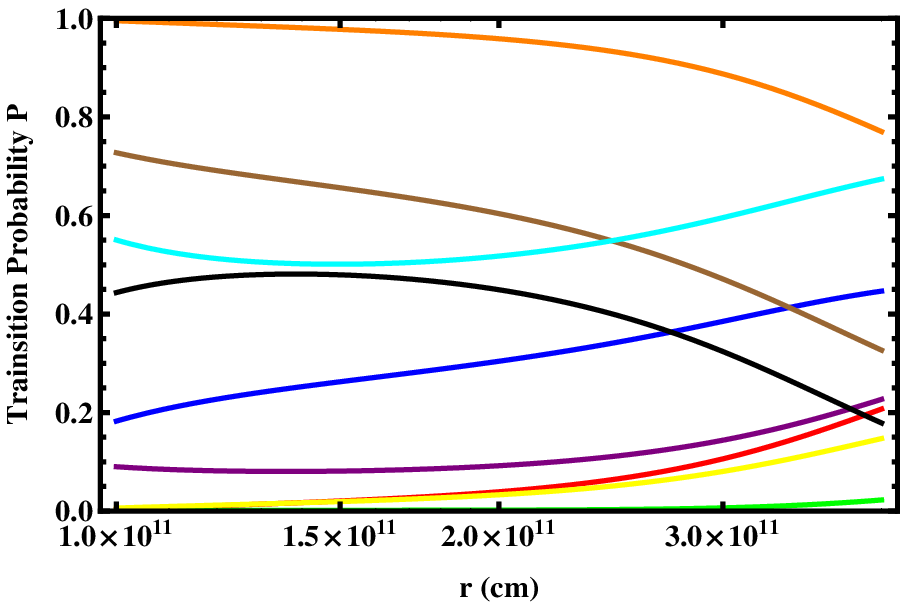}
\label{fig4:subfig:c}
\end{minipage}
}
\subfigure[Transition probabilities evolve versus radius for $1\rm PeV$ neutrinos]{
\begin{minipage}[b]{0.45\textwidth}
\includegraphics[width=1\textwidth]{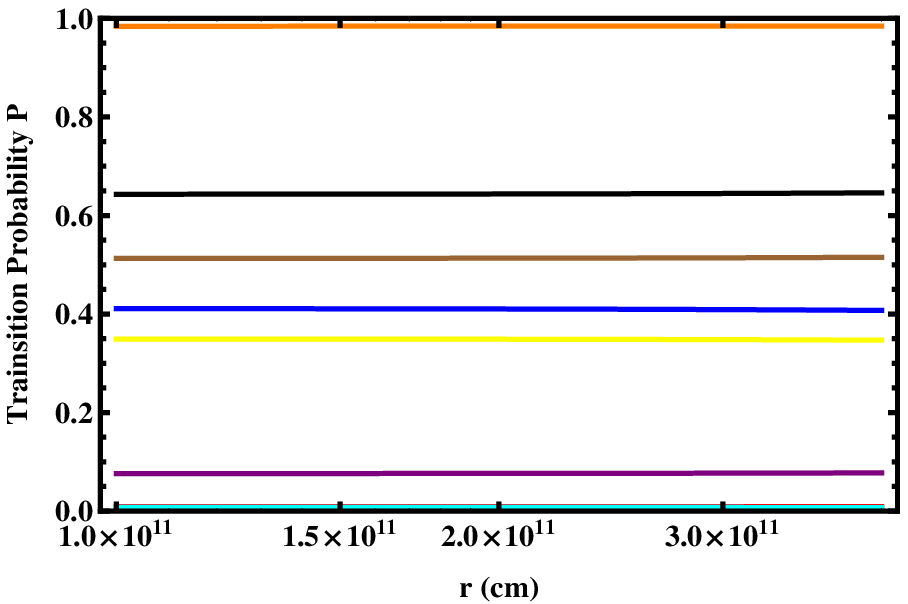}
\label{fig4:subfig:d}
\end{minipage}
}
\caption{Transition probabilities in matter for neutrinos with different energies at different radii. Nine different colors represent nine mutual transition probabilities and are listed below: $P_{ee}\to$ red, $P_{e\mu}\to$ orange, $P_{e\tau}\to$ green, $P_{\mu e}\to$ blue, $P_{\mu\mu}\to$ brown, $P_{\mu\tau}\to$ purple, $P_{\tau e}\to$ black, $P_{\tau\mu}\to$ yellow, $P_{\tau\tau}\to$ cyan. These apply for all four subfigures above. \label{Figure 4}}
\end{figure}
\begin{figure}
\centering
\subfigure[Flavor ratio evolves versus radius for $1\rm TeV$ neutrinos]{
\begin{minipage}[b]{0.45\textwidth}
\includegraphics[width=1\textwidth]{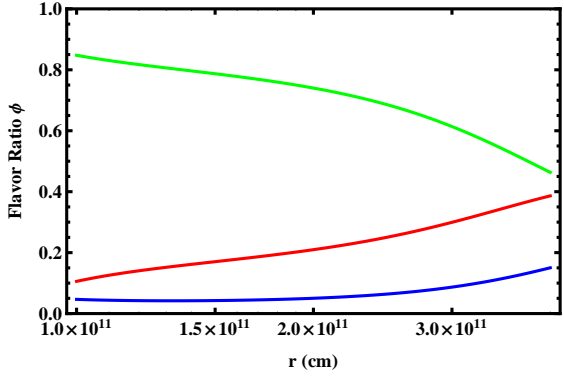}
\label{fig5:subfig:a}
\end{minipage}
}
\subfigure[Flavor ratio evolves versus radius for $10\rm TeV$ neutrinos]{
\begin{minipage}[b]{0.45\textwidth}
\includegraphics[width=1\textwidth]{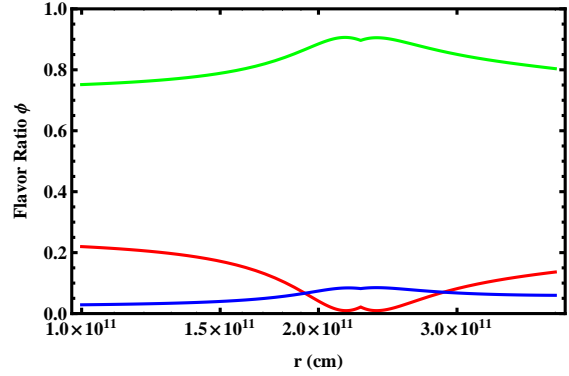}
\label{fig5:subfig:b}
\end{minipage}
}
\subfigure[Flavor ratio evolves versus radius for $100\rm TeV$ neutrinos]{
\begin{minipage}[b]{0.45\textwidth}
\includegraphics[width=1\textwidth]{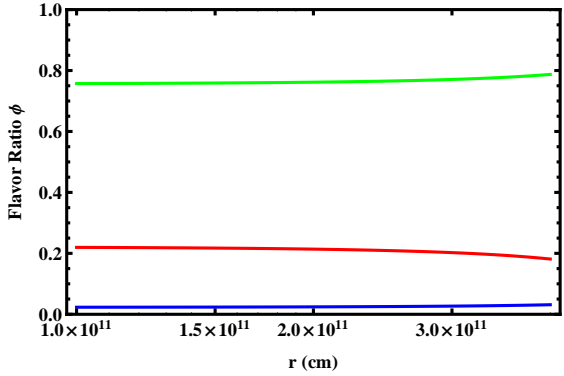}
\label{fig5:subfig:c}
\end{minipage}
}
\subfigure[Flavor ratio evolves versus radius for $1\rm PeV$ neutrinos]{
\begin{minipage}[b]{0.45\textwidth}
\includegraphics[width=1\textwidth]{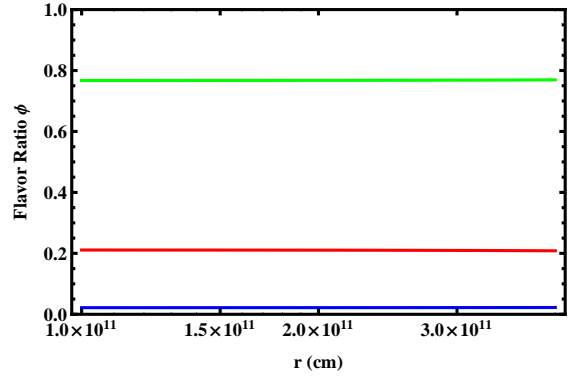}
\label{fig5:subfig:d}
\end{minipage}
}
\caption{Flavor ratio in matter evolves versus radius for neutrinos with four different energies. In all four subfigures above, the red line represents the fraction of $\nu_e$, the green line is $\nu_\mu$ and the blue line is $\nu_\tau$. \label{Figure 5}}
\end{figure}

We plot the nine mutual transition probabilities between three neutrino flavors as functions of neutrino energy at different radii in Figure~\ref{fig4:subfig:a}-~\ref{fig4:subfig:b} and the evolution with propagation distance for given energies in Figure~\ref{fig4:subfig:c}-~\ref{fig4:subfig:d}. Given an initial flavor ratio $\phi_{\nu_e}^0:\phi_{\nu_\mu}^0:\phi_{\nu_\tau}^0=1:2:0$, we can plot how the flavor ratio changes with energy and distance when neutrinos are still in the envelope in Figure~\ref{fig5:subfig:a}-~\ref{fig5:subfig:d}. The deviation from $1:2:0$ at born site $r_0$ is due to level crossing effect for different energies. We can see the trend that the flavor ratio changes more gently for neutrinos with higher energy. For $\rm PeV$ neutrinos, the flavor ratio is almost a constant value $\phi_{\nu_e}:\phi_{\nu_\mu}:\phi_{\nu_\tau}\simeq0.21:0.77:0.02$ in the envelope.
\subsubsection{Neutrino Oscillation from Progenitor to Earth}
\begin{figure}
\centering
\subfigure[Flavor ratio versus energy when neutrinos are just leaving the progenitor]{
\begin{minipage}[b]{0.45\textwidth}
\includegraphics[width=1\textwidth]{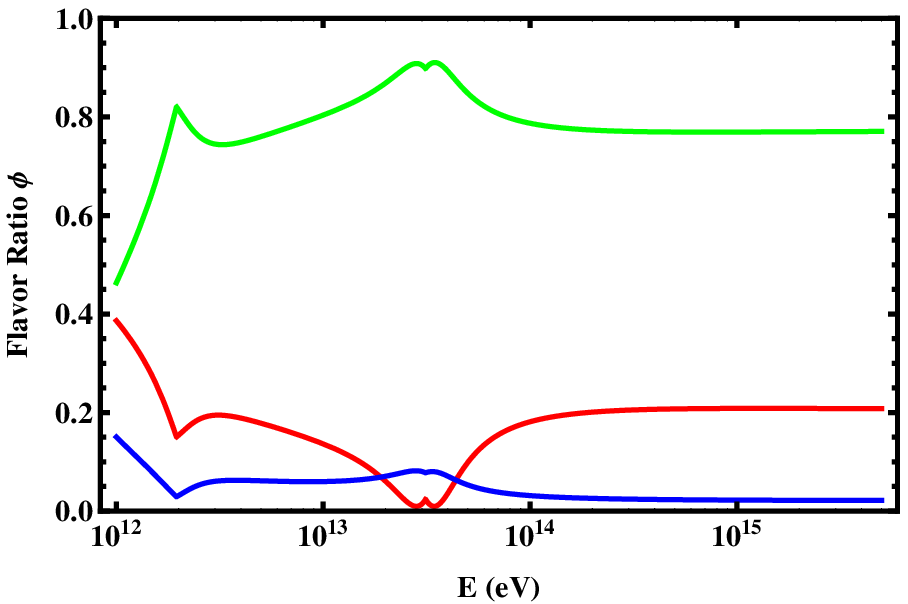}
\label{fig6:subfig:a}
\end{minipage}
}
\subfigure[Flavor ratio versus energy on Earth]{
\begin{minipage}[b]{0.45\textwidth}
\includegraphics[width=1\textwidth]{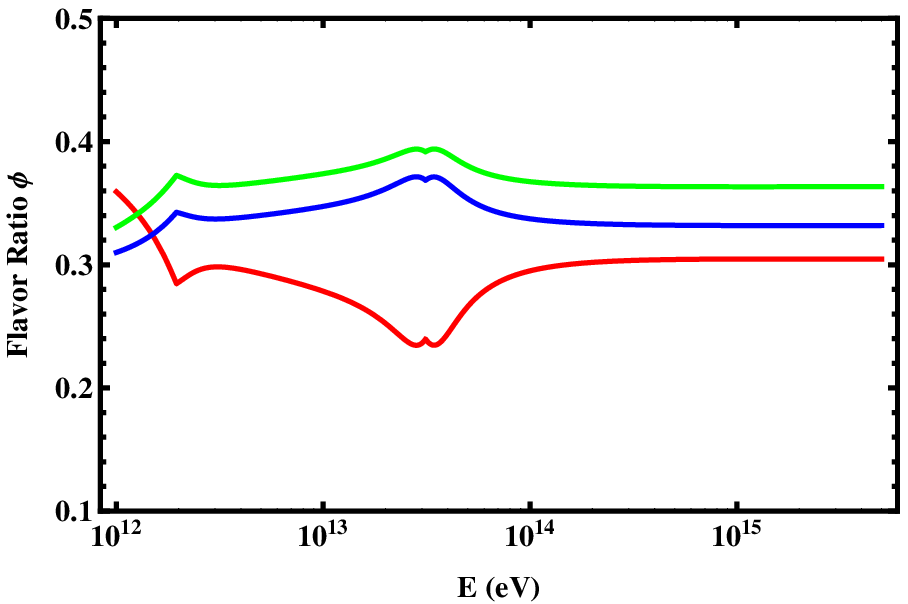}
\label{fig6:subfig:b}
\end{minipage}
}
\caption{Neutrino flavor ratio versus energy before and after long distance vacuum oscillation. Same as Figure~\ref{Figure 5}, the red line represents the fraction of $\nu_e$, the green line is $\nu_\mu$ and the blue line is $\nu_\tau$. \label{Figure 7}}
\end{figure}

Neutrinos go through vacuum oscillation after leaving the progenitor surface and has been well understood. We can express the transition probability $P_{\alpha\beta}^0$ as the first-order expansion of the small parameter $\sin{\theta_{13}}$ \citep{xin06}:
$$P_{ee}^0=1-\frac{1}{2}\sin^2{2\theta_{12}}\,,$$
$$P_{e\mu}^0=P_{\mu e}^0=\frac{1}{2}\sin^2{2\theta_{12}}\cos^2{\theta_{23}}+\frac{1}{4}\sin4\theta_{12}\sin2\theta_{23}\sin\theta_{13}\cos\delta\,,$$
$$P_{e\tau}^0=P_{\tau e}^0=\frac{1}{2}\sin^2{2\theta_{12}}\sin^2{\theta_{23}}-\frac{1}{4}\sin4\theta_{12}\sin2\theta_{23}\sin\theta_{13}\cos\delta\,,$$
$$P_{\mu\mu}^0=1-\frac{1}{2}\sin^2{2\theta_{23}}-\frac{1}{2}\sin^2{2\theta_{12}}\cos^4\theta_{23}-\frac{1}{2}\sin4\theta_{12}\sin2\theta_{23}\cos^2\theta_{23}\sin\theta_{13}\cos\delta\,,$$
$$P_{\mu\tau}^0=P_{\tau\mu}^0=\frac{1}{2}\sin^2{2\theta_{23}}-\frac{1}{8}\sin^2{2\theta_{12}}\sin^2{2\theta_{23}}+\frac{1}{8}\sin4\theta_{12}\sin4\theta_{23}\sin\theta_{13}\cos\delta\,,$$
$$P_{\tau\tau}^0=1-\frac{1}{2}\sin^2{2\theta_{23}}-\frac{1}{2}\sin^2{2\theta_{12}}\sin^4{\theta_{23}}+\frac{1}{2}\sin4\theta_{12}\sin2\theta_{23}\sin^2\theta_{23}\sin\theta_{13}\cos\delta\,. \eqno (10)$$
The flavor ratio on Earth is
\begin{displaymath}
\left( \begin{array}{ccc}
\phi_{\nu_e} \\
\phi_{\nu_\mu} \\
\phi_{\nu_\tau}
\end{array} \right)_{\rm Earth}=\left( \begin{array}{ccc}
P_{ee}^0 & P_{e\mu}^0 & P_{e\tau}^0\\
P_{\mu e}^0 & P_{\mu\mu}^0 & P_{\mu\tau}^0\\
P_{\tau e}^0 & P_{\tau\mu}^0 & P_{\tau\tau}^0
\end{array} \right)\left( \begin{array}{ccc}
\phi_{\nu_e} \\
\phi_{\nu_\mu} \\
\phi_{\nu_\tau}
\end{array} \right)_{\rm source}.
\eqno (11)
\end{displaymath}
We plot the flavor ratio versus neutrino energy on just leaving the progenitor and on Earth in Figure~\ref{fig6:subfig:a}-~\ref{fig6:subfig:b}. We can see clearly that the flavor ratio varies with energy in the range of less than $100\rm TeV$, while the flavor ratio keeps invariant  $\phi_{\nu_e}:\phi_{\nu_\mu}:\phi_{\nu_\tau}\simeq0.30:0.37:0.33$ for $\rm PeV$ neutrinos.

\subsection{Dependence on $\alpha$}
It is reasonable to argue that the final neutrino flavor ratio depends on the density profile of the progenitor envelope. Apparently, with the same assumed envelope mass and radius, different values of the power law index lead to different ambient envelope densities, thus effective potentials are different. This will have an impact on the resonance conditions, effective mixing angles in matter and transition probabilities. In this subsection, we would like to investigate how large this impact could be. Here we adopt the same helium progenitor but with different power law index $\alpha=2.5,\,2.7$. Respectively, we can write them as $\rho(r)=2.52\times10^{26}r^{-2.5}\,\rm g\,cm^{-3}$ and $\rho(r)=3.16\times10^{28}r^{-2.7}\,\rm g\,cm^{-3}$ and all calculations have been repeated for these two cases.
\par
\begin{figure}
\centering
\subfigure[Comparison of flavor ratios versus energy when neutrinos are just leaving the progenitors for three different envelope power law indexes]{
\begin{minipage}[b]{0.45\textwidth}
\includegraphics[width=1\textwidth]{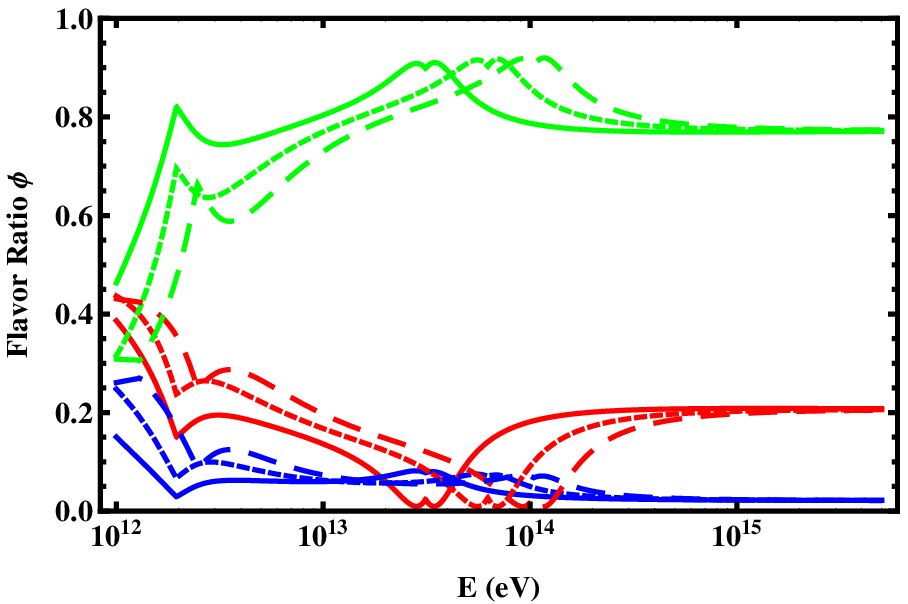}
\label{fig7:subfig:a}
\end{minipage}
}
\subfigure[Comparison of flavor ratios versus energy on Earth for three different progenitor envelope power law indexes]{
\begin{minipage}[b]{0.45\textwidth}
\includegraphics[width=1\textwidth]{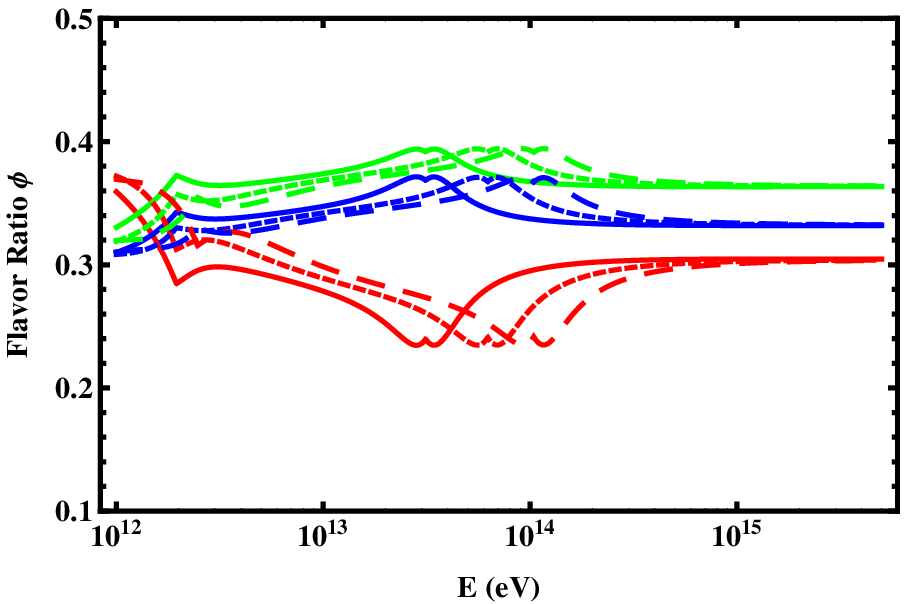}
\label{fig7:subfig:b}
\end{minipage}
}
\caption{The dependence of flavor ratios versus energy before and after long distance vacuum oscillation on progenitor envelope power law index. The red lines represent the fractions of $\nu_e$, the green lines are $\nu_\mu$ and the blue lines are $\nu_\tau$. Besides, solid lines are responsible for $\alpha=2$, dotted lines are for $\alpha=2.5$ and dashed lines are for $\alpha=2.7$. \label{Figure 7}}
\end{figure}

We present our result in Figure~\ref{Figure 7}. For simplicity, we only show the flavor ratio at the surface of the progenitor (Figure~\ref{fig7:subfig:a}) and on Earth ( Figure~\ref{fig7:subfig:b}). It is clear that $\alpha$ has an impact on the flavor ratio. At a given radius, the resonance conditions are shifted to higher energies for larger $\alpha$. The lower limit of neutrino energy for constant flavor ratio is highest for $\alpha=2.7$, which is several $\rm PeV$, compared with sub$\rm PeV$ for $\alpha=2.5$ and $\sim100\rm TeV$ for $\alpha=2$. The reason is that the effective potential of neutrinos is lower for a steeper envelope density profile at the same radius, so higher neutrino energies are required to reach resonance conditions. Furthermore, the flavor ratio is evidently different before it reaches constant for the three cases, so that we can use the observed ratio of neutrino energy from $\rm TeV$ to several hundred $\rm TeVs$ to probe the stellar structure, provided that we can observe precursor neutrinos of a GRB with $\rm km^2$ scale detectors like IceCube in the future.

\section{Discussions and Conclusions}
High-energy neutrinos can be produced while the jet is still propagating in the envelope and LL-GRBs with typical parameters are responsible for $\rm TeV-PeV$ neutrinos. These neutrinos will oscillate with matter effect in the envelope and go through vacuum oscillation after leaving the progenitor till they arrive at the Earth. We investigate the three-neutrino mixing properties with matter effect and then get an expected flavor ratio on Earth, given an initial ratio $\phi_{\nu_e}^0:\phi_{\nu_\mu}^0:\phi_{\nu_\tau}^0=1:2:0$.
\par
We notice that adiabatic conversion is violated because level crossing effect is non-negligible for such high energy neutrinos. After calibrating this effect, we get the neutrino mixing angles in matter and nine transition probabilities. We find that the effective mixing angles tend to be zero for neutrinos on the high energy end ($\sim \rm PeV$), resulting in constant transition probability and constant flavor ratio. For $\rm PeV$ neutrinos, we always get $\phi_{\nu_e}:\phi_{\nu_\mu}:\phi_{\nu_\tau}\simeq0.30:0.37:0.33$ on Earth.
\par
From our expectations, the final neutrino ratio will depend on the density profile parameter $\alpha$. We take $\alpha=2,\,2.5,\,2.7$ to verify the dependence and the result is clear in Figure~\ref{Figure 7}. While the flavor ratio on the high energy end keeps constant, it is evidently different for neutrinos with energy from $\rm TeV$ to several hundred $\rm TeVs$ , thus providing a promising way to probe the stellar structure in the future.
\par
In this paper, we only investigate the high-energy neutrino oscillation of one typical LL-GRB for the given parameters. Changing these parameters may have an impact on the flavor ratio-neutrino energy dependence in TeV range: at a given radius, the resonance energy may differ and the lower limit of neutrino energy for constant flavor ratio is also different, similar to the features shown in Figure~\ref{Figure 7}. However, it does not influence the final flavor ratio of PeV neutrinos. That is because the envelope is always too ``dense" for PeV neutrinos, the effective mixing angles of which always tend to be constant zero. The case is the same when changing envelope temperature $T$. Though the number density of electrons varies for different $T$, the effective mixing angles for PeV neutrinos are always zero and constant the flavor ratio $\phi_{\nu_e}:\phi_{\nu_\mu}:\phi_{\nu_\tau}\simeq0.30:0.37:0.33$ on Earth is still in expectation. However, this constant value may differ from $0.30:0.37:0.33$ due to the uncertainties of vacuum oscillation parameters and neutrino mass hierarchy.
\par
The IceCube experiment has recently reported the observation of 37 high-energy ($\ge 30 \rm TeV$) neutrino events, separated into 28 showers and 9 muon tracks, being consistent with an extraterrestrial origin \citep{ice14}. To correlate with the observation, we take the hypothesis that all the IceCube neutrinos are produced by LL-GRBs before jet-breakout. \citet{men14} claimed that $\phi_{\nu_e}:\phi_{\nu_\mu}:\phi_{\nu_\tau}=1:1:1$ at Earth is disfavored at $92\%$ C.L. with the recently released 3-yr data. On one hand, neutrino flavor ratio in the detector may be changed by the Earth matter effect \citep{var14}. On the other hand, this does not conflict with our conclusion. We just recommend to do an analysis of observed shower-to-track ratio in different energy bins when doing data reduction since we know that the flavor ratio depends strongly on neutrino energy and an overall $1:1:1$ ratio does not make any sense. Nevertheless, we expect constant flavor ratio for $\rm PeV$ neutrinos but we have observed only three $\rm PeV$ events now. So our result is to be verified with a larger dataset of $\rm TeV-PeV$ neutrinos of IceCube in the future. If this constant value appears in next few decades, when we would have observed tens of PeV neutrino events, we can constrain the structure of LL-GRB progenitors and vacuum oscillation parameters by exactly measuring this value. If there is no sign of such a constant ratio, the most probable reason is that there exist other dominate PeV neutrino sources. The hypothesis that all observed neutrinos are produced in the jet propagation process of LL-GRBs may be not so complete because they may also origin from other cosmic-ray sources like AGNs or they can be produced in other stages of a GRB event such as by internal shocks and external shocks. Especially, we hope that one day we could observe the neutrino precursor of a GRB event, this neutrino-GRB correlation is crucial for our understanding about the structure of the progenitor envelope and the jet propagation dynamics. For a complete comparison with the observation, the flavor ratio in TeV range of diffuse neutrino background produced by GRBs needs to be done and is beyond the scope of this paper.
\par
\acknowledgments
We thank an anonymous referee for his/her helpful suggestions. This work is supported by the National Basic
Research Program of China (973 Program, grant 2014CB845800) and the
National Natural Science Foundation of China (grant 11033002).

\clearpage

\begin{thebibliography}{}

\bibitem[Ando \& Beacom(2005)]{and05} Ando, S., Beacom, J. F. 2005, \prl, 95, 061103
\bibitem[Alvarez-Muniz et al.(2000)]{alv00} Alvarez-Muniz, J., Halzen, F., \& Hooper, D. W. 2000, \prd, 62, 093015
\bibitem[Bahcall \& Waxman(2001)]{bah01} Bahcall, J., \& Waxan, E. 2001, \prd, 64, 023002
\bibitem[Becker(2008)]{bec08} Becker, J. K. 2008, Phys. Rep., 458, 173
\bibitem[Chen(1985)]{che85} Chen, H. H. 1985, \prl, 55, 1534
\bibitem[Dai \& Lu(2001)]{dai01} Dai, Z. G., \& Lu, T. 2001, \apj, 551, 249
\bibitem[Dermer et al.(2003)]{der03} Dermer, C. D., \& Atoyan, A. 2003, \prl, 91, 071102
\bibitem[Dighe \& Smirnov(2000)]{dig00} Dighe, A. S., \& Smirnov, A. Y. 2000, \prd, 62, 033007
\bibitem[Enberg et al.(2009)]{enb09} Enberg, R., Reno, M. H., \& Sarcevic, I. 2009, \prd, 79, 053006
\bibitem[Fogli et al.(2012)]{fog12} Fogli, G. L., Lisi, E., Marrone, A., Montanino, D., Palazzo, A., \& Rotunno, A. M. 2012, \prd, 86, 013012
\bibitem[Fraija(2014)]{fra14} Fraija, N. 2014, \mnras, 437, 2187
\bibitem[Gao et al.(2012)] {gao12} Gao, S., Asano, K., \& M\'esz\'aros, P. 2012, JCAP, 1211, 058
\bibitem[Guetta \& Granot(2003)]{gue03} Guetta, D., \& Granot, J. 2003, \prl, 90, 201103
\bibitem[Guetta et al.(2004)]{gue04} Guetta, D., Hooper, D., Alvarez-Mu\~niz, J., Halzen, F., \& Reuveni, E. 2004, Astroparticle Physics, 20, 429
\bibitem[Horiuchi \& Ando(2008)]{hor08} Horiuchi, S., \& Ando, S. 2008, \prd, 77, 063007
\bibitem[IceCube Collaboration(2014)]{ice14} IceCube Collaboration, 2014, \prl, 113, 101101
\bibitem[Kashti \& Waxman(2005)]{kas05} Kashti, T., \& Waxman, E. 2005, \prl, 95, 181101
\bibitem[Katz et al.(2010)]{kat10} Katz, B., Budnik, R., \& Waxman, E. 2010, \apj, 716, 781
\bibitem[Kuo(1989)]{kuo89} Kuo, T. K. 1989, \prd, 39, 1930
\bibitem[Levinson \& Bromberg(2008)]{lev08} Levinson, A., \& Bromberg, O. 2008, \prl, 100, 131101
\bibitem[Li et al.(2002)]{li02} Li, Z., Dai, Z. G., \& Lu, T. 2002, \aap, 396, 303
\bibitem[Lund \& Kneller(2013)]{lun13} Lund, T., \& Kneller J. P. 2013, \prd, 88, 3008
\bibitem[MacFadyen \& Woosley(1999)]{mac99} MacFadyen, A., \& Woosley, S. E. 1999, \apj, 524, 262
\bibitem[Mena et al.(2007)]{men07}Mena, O., Mocioiu, I., \& Razzaque, S. 2007, \prd, 75, 063003
\bibitem[Mena et al.(2014)]{men14}Mena, O., Palomares-Ruiz, S., \& Vincent, A. C. 2014, \prl, 113, 1103
\bibitem[M\'esz\'aros \& Rees(2000)]{mes00} M\'esz\'aros, P., \& Rees, M. J. 2000, \apj, 541, L5-L8
\bibitem[M\'esz\'aros \& Waxman(2001)]{mes01} M\'esz\'aros, P., \& Waxman, E. 2001, \prl, 87, 171102
\bibitem[M\'esz\'aros(2006)]{mes06} M\'esz\'aros, P. 2006, Rept. Prog. Phys., 69, 2259
\bibitem[Mikheyev \& Smirnov(1985)]{mik85} Mikheyev, S. P., \& Smirnov, A. Y. 1985, Yad. Fiz., 42, 1441
\bibitem[Murase et al.(2006)]{mura06} Murase, K., Ioka, K., Nagataki, S., \&Nakamura, T. 2006, \apj, 651, L5
\bibitem[Murase \& Nagataki(2006)]{mur06} Murase, K., \& Nagataki, S. 2006, \prd, 73, 063002
\bibitem[Murase(2008)]{mur08} Murase, K. 2008, \prd, 78, 101302
\bibitem[Murase \& Ioka(2013)]{mur13} Murase, K., \& Ioka, K. 2013, \prl, 111, 121102
\bibitem[Osorio Oliveros et al.(2013)]{oli13} Osorio Oliveros, A. F., Sahu, S., \& Sanabria, J. C. 2013, The European Physical Journal C, 73, 2574
\bibitem[Piran(2005)]{pir05} Piran, T. 2005, Rev. Mod. Phys., 76, 1143
\bibitem[Pruet(2003)]{pru03} Pruet, J. 2003, \apj, 591, 1104-1109
\bibitem[Rachen \& M\'esz\'aros(1998)]{rac98} Rachen, J. P., \& M\'esz\'aros, P. 1998, \prd, 58, 123005
\bibitem[Razzaque et al.(2003)]{raz03} Razzaque, S., M\'esz\'aros, P., \& Waxman, E. 2003, \prl, 90, 241103
\bibitem[Razzaque \& Smirnov(2010)]{raz10} Razzaque, S., \& Smirnov, A. Y. 2010, Journal of High Energy Physics, 3, 31
\bibitem[Razzaque(2013)]{raz13} Razzaque, S. 2013, \prd, 88, 103003
\bibitem[Rees \& M\'esz\'aros(2005)]{ree05} Rees, M. J., \& M\'esz\'aros, P. 2005, \apj, 628, 847
\bibitem[Sahu \& Zhang(2010)]{sah10} Sahu, S., \& Zhang, B. 2010, Research in Astronomy and Astrophysics, 10, 943
\bibitem[Scholberg(2012)]{sch12} Scholberg, K. 2012, Annual Review of Nuclear and Particle Science, 62, 81
\bibitem[Varela et al.(2014)]{var14} Varela, K., Sahu, S., Osorio Oliveros, A. F., \& Sanabria, J. C. 2014 \href{http://arxiv.org/abs/1411.7992}{arXiv:1411.7992}.
\bibitem[Vietri(1995)]{vie95} Vietri, M. 1995, \apj, 453, 883
\bibitem[Wang \& Dai(2009)]{wan09} Wang, X. Y., \& Dai, Z. G. 2009, \apj, 691, 67
\bibitem[Waxman \& Bahcall(1997)]{wax97} Waxman, E., \& Bahcall, J. 1997, \prl, 78, 2292
\bibitem[Waxman \& Bahcall(1999)]{wax99} Waxman, E., \& Bahcall, J. 1998, \prd, 59, 023002
\bibitem[Waxman \& Bahcall(2000)]{wax00} Waxman, E., \& Bahcall, J. 2000, \apj, 541, 707
\bibitem[Wolfenstein(1978)]{wol78} Wolfenstein, L. 1978, \prd, 17, 2369
\bibitem[Woosley(1993)]{woo93} Woosley, S. E. 1993, \apj, 405, 273
\bibitem[Xiao \& Dai(2014)]{xia14} Xiao, D., \& Dai, Z. G. 2014, \apj, 790, 59
\bibitem[Xing \& Zhou(2006)]{xin06} Xing, Z. Z., \& Zhou, S. 2006, \prd, 74, 3010
\bibitem[Yasuda(2014)]{yas14} Yasuda, O. 2014, \prd, 89, 3023
\bibitem[Zhang \& M\'esz\'aros(2004)]{zha04} Zhang, B., \& M\'esz\'aros, P. 2004, Int. J. Mod. Phys. A 19, 2385

\end{thebibliography}
\end{document}